\documentclass[pra,twocolumn,showpacs]{revtex4}
\usepackage{times}
\usepackage{amsbsy,amssymb}
\usepackage{graphicx,color}
\input{Qcircuit}

\newcommand{\avg}[1]{\langle #1 \rangle}

\def\one{{\mathchoice {\rm 1\mskip-4mu l} {\rm 1\mskip-4mu l} {\rm
1\mskip-4.5mu l} {\rm 1\mskip-5mu l}}}

\newcommand{\trace}{\mbox{tr}}

\newcommand{\ketbra}[2]{ \ket{#1}_{\! #2}^{#2} \!\! \bra{#1}}
\newcommand{\ketbras}[3]{ \ket{#1}_{\! #3}^{#3} \!\! \bra{#2}}

\newcommand{\controlu}[1]{^{c\ }\!\!#1}

\providecommand{\ignore}[1]{}

\def\openone{\leavevmode\hbox{\small1\kern-3.8pt\normalsize1}}
\def\RR{{\rm I\kern-.2emR}}
\def\tr{{\rm tr}  \,}

\def\fh{\mathfrak{h}}

\def\fsu{\mathfrak{su}}

\def\openone{\leavevmode\hbox{\small1\kern-3.8pt\normalsize1}}
\def\RR{{\rm I\kern-.2emR}}

\def\cO{{\cal O}}

\def\cN{ {\cal {N}} }

\providecommand{\ignore}[1]{}

\newcommand{\shortqph}[1]{}

\begin{document}
\title{Parameter Estimation with Mixed-State Quantum Computation}

\author{Sergio Boixo}
\email{boixo@unm.edu}

\affiliation{Los Alamos National Laboratory, Los Alamos, NM 87545, USA}
\affiliation{University of New Mexico, Albuquerque , NM 87131, USA}

\author{Rolando D. Somma}
\email{somma@lanl.gov}

\affiliation{Los Alamos National Laboratory, Los Alamos, NM 87545, USA}

\date{\today}

\begin{abstract}
  We present a quantum algorithm to estimate parameters at the quantum
  metrology limit using deterministic quantum computation with one
  bit.  When the interactions occurring in a quantum system are
  described by a Hamiltonian $H= \theta H_0$, we estimate $\theta$ by
  zooming in on previous estimations and by implementing an adaptive
  Bayesian procedure. The final result of the algorithm is an updated
  estimation of $\theta$ whose variance has been decreased in
  proportion to the time of evolution under $H$.  For the problem of
  estimating several parameters, we implement dynamical-decoupling
  techniques and use the results of single parameter estimation. The
  cases of discrete-time evolution and reference frame alignment are
  also studied within the adaptive approach.
\end{abstract}


\maketitle

%%%%%%%%%%%%%%%%%%%%%%%%%%%%%%%%%%%%%%%%
%%%%%%%%%%%%%%%%%%%%%%%%%%%%%%%%%%%%%%%%
\section{Introduction}
\label{intro}
Quantum mechanics provides new resources that allow us to determine
physical properties at the highest possible accuracy established by
generalized uncertainty relations~\cite{Holevo, Helstrom, BCM96}.
Exploiting quantum coherence enables us to estimate
parameters~\cite{GLM04, VARIOUS1} and expectation values of
observables~\cite{KOS07} with better resource scaling than classically
possible. In this paper, we are interested in the estimation of
interaction parameters (e.g., external fields), when the interaction
acts independently on $n$ quantum subsystems in a
probe~\cite{comment-0}.  We quantify our  resource of interest 
by  $N=nT$, given by the product of the number of
subsystems and the interaction time $T$.  The standard quantum
limit (SQL) precision in the estimation of an interaction parameter is
of order $\cO(1/\sqrt N)$, achievable with $\cO(N)$ independent
measurements at a fixed $T$.  The optimal precision for
such an estimation, however, is given by the Heisenberg limit and is
known to be of order $\cO(1/N)$. Achieving it requires the preparation
of entangled quantum state in the probe~\cite{GLM04}.

We are interested in estimating parameters at the so called quantum
metrology limit (QML). This can be obtained by a series of estimations performed
at different interaction times, while keeping the size of the probe
fixed~\cite{VARIOUS1,KOS07}. If for a total interaction time $T$, the
precision of the estimation is of order $\cO(1/T)$, we say that the
QML has been achieved.  This \emph{sequential} protocol, which is the
one exploited in this paper, does not require quantum entanglement in
the input state, although the response to uncorrelated decoherence,
for an unconstrained interaction time, is the same as that of the
entangled protocol~\cite{decoh}.  Any method that allows us
to achieve the QML provides clearly an improvement over the SQL, since
for the same amount of resources (i.e., $N=nT$), the returned precision can be
highly enhanced.

Quantum methods (algorithms) designed to beat the SQL could have a
wide range of applications, from highly sensitive
magnetometry~\cite{GSM04} to atomic clock
synchronization~\cite{BIW96}. In addition, phase estimation, a problem
related to parameter estimation, is one of the cornerstones of Quantum
Computation~\cite{CEMM98, WZ06}. In this paper we show
that the QML can be achieved in some cases even if the initial state
is the completely mixed state of all except one of the quantum
systems, avoiding the complexity associated with initial pure, entangled, state
preparation.  Although here we consider multi-qubit probes,
generalization of our algorithms to higher dimensional systems is
straightforward.

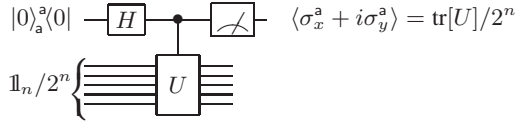
\begin{figure}[!h]
\begin{equation*}
\hspace{-2.4cm}
  \Qcircuit @C=.5em @R=-.5em {
     \lstick{\ketbra{0}{\sf a}}  &\qw & \gate{H} & \ctrl{1} & \meter  & \qw & \rstick{\langle \sigma_x^{\sf a} + i \sigma_y^{\sf a} \rangle=\trace[U]/2^{n}} & \push{\rule{0em}{4em}} \\
    &  \qw & \qw &  \multigate{4}{U} & \qw &  \\
    &  \qw & \qw &  \ghost{U} & \qw & \\
    \lstick{\mbox{$\one_{n}/2^{n}$}}  & \qw  &\qw &\ghost{U} & \qw&  \\
    &   \qw & \qw  &  \ghost{U} & \qw   &\\
    &   \qw & \qw  &  \ghost{U} & \qw   \gategroup{2}{2}{6}{1}{.5em}{\{} 
  }
\end{equation*}
\caption{DQC1 circuit for estimating the trace of a unitary $U$. The
  filled circle denotes that $U$ acts on the probe when the state of
  the ancilla (control qubit) is $\ket {1}_{\sf a}$, and $H$ is the
  Hadamard gate. $\langle \sigma_x^{\sf a} \rangle$ and $\langle
  \sigma_y^{\sf a} \rangle$ are the expectation values of the
  corresponding Pauli operators on the ancilla $\sf
  a$.}\label{dqc1fig}
\end{figure}

Specifically, we use deterministic quantum computation with one bit
(DQC1), which was initially described in Ref.~\cite{KL98} in the
context of high temperature ensemble quantum computation using
liquid-state NMR techniques~\cite{LKC02}.  Although less powerful than
the standard model of quantum computation, DQC1 is believed to
outperform the classical probabilistic computational
model~\cite{VARIOUS2}.  In DQC1, the initial state $\rho_0$ of a set
of $n+1$ qubits corresponds to having the first (ancilla) qubit $\sf
a$ in the pure state $\ket{0}_{\sf a}$, while the state of the
remaining $n$ qubits (probe) is completely mixed. That is,
\begin{equation}
\label{dqc1is}
\rho_0 = \frac{1}{2^{n}} (\ketbra{0}{\sf a} \otimes \one_{n})\;.
\end{equation}
The state $\rho_0$ is then unitarily evolved and DQC1 returns a noisy
expectation value of a Pauli operator on the ancilla qubit.  If the
evolution is performed by applying a unitary operation controlled by
${\sf a}$ (i.e., a controlled-$U$ or $\controlu U$ operation), DQC1
allows us to estimate the renormalized trace of $U$ at a certain,
fixed precision (Fig.~\ref{dqc1fig}).

We assume that a single run of the DQC1 algorithm returns an unbiased
renormalized trace estimation with normal distribution $ \cN(
\trace[U]/2^{n} , \Delta^2 )$ of standard deviation
$\Delta$~\cite{comment-1}.  Of course, $\Delta $ can be reduced by a
factor of $\sqrt{K}$ (i.e., SQL) if the algorithm of
Fig.~\ref{dqc1fig} is repeated $K$ times. In fact, this is the
situation in NMR where repetition reduces the signal-to-noise ratio
(SNR) at the SQL. 

Consider now, for example, the typical case where an unknown external
magnetic field interacts with the $n$ qubits of the probe
(Fig.~\ref{dqc1fig}), determining an interaction Hamiltonian of the
form
\begin{equation}
\label{mfield}
H'= \theta \sum_{j=1}^n \sigma_z^j\;.
\end{equation}
We seek to estimate $\theta$. When the probe interacts with
the field for  time $T$, the $n$-qubit state is evolved by applying
the corresponding (unitary) evolution operator
$W'(T)=e^{-iH'T}$. Replacing $\controlu U$ by $\controlu W'(T)$ in
Fig.~\ref{dqc1fig}, the final $(n+1)-$qubit state right before the
measurement is
\begin{eqnarray}
\rho_f = \frac{1}{2^{n+1}}[\ketbra{0}{\sf a} \otimes \one_n + \ketbras{1}{0}{\sf a}  \otimes W'(T) + \\
\nonumber
\ketbras{0}{1}{\sf a}  \otimes W'^\dagger(T)+ \ketbra{1}{\sf a}  \otimes \one_n]\;.
\end{eqnarray}
Using the trace properties of the Pauli operators, we obtain $\langle
\sigma_x^{\sf a}\rangle = \tr[\rho_f \sigma_x^{\sf a}] = (\cos(\theta
T))^n$.  If $\Delta_x>0$ denotes the standard deviation in the
estimation of $\avg{ \sigma_x^{\sf a}}$, a first approximation error
formula determines
\begin{equation}
\label{deltatheta}
\Delta_\theta \approx \frac{\Delta_x}{|\partial \langle\sigma_x^{\sf a}\rangle / \partial \theta|}\;,
\end{equation}
with $\Delta_\theta$ the uncertainty in the estimation of $\theta$.

Let $\ket +$ denote the single qubit state $(\ket 0 + \ket 1)/\sqrt 2$. The output
signal of the previous algorithm is $|\bra{+_1 \ldots +_{n/2}}W'(T)\ket{+_1 \ldots +_{n/2}}|^2$,
which is the probability of
measuring $\ket{+_1 \ldots +_{n/2}}$ after its evolution under $W'(T)$.
Thus, the output precision of
Eq.~(\ref{deltatheta}) is upper bounded by $\Delta_\theta
=\cO[1/(\sqrt{n} T)]$, and the Heisenberg limit is not achieved
when scaling $n$. Nevertheless, for fixed (small) $n$, we obtain $\Delta_\theta
=\cO(1/T)$, yielding the QML.

The previous estimation method has some important disadvantages. The first
one concerns the use of the controlled $\controlu W'(T)$ operation
which, due to technological difficulties or to the nature of the
problem, may be impossible. In fact, this is the case in reference
frame alignment, as we discuss in Sec.~\ref{rfa}.  Second, it is clear
that $(\cos(\theta T))^n$ approaches $0$ exponentially with $n$ for $T
\theta \ne p \pi$, which is usually the case, as $\theta$ is unknown.
As a consequence, the SNR of the outcome is weakened, especially for
$n \gg 1$. 

%  For fixed $n$ we
% obtain $\Delta_\theta = \cO(1/T)$, so the QML can be achieved by
% scaling $T$~\cite{comment-1b}.

% However, this method has some important disadvantages. The first
% one concerns the use of the controlled $\controlu W'(T)$ operation
% which, due to technological difficulties or to the nature of the
% problem, may be impossible. In fact, this is the case in reference
% frame alignment, as we discuss in Sec.~\ref{rfa}.  Second, it is clear
% that $(\cos(\theta T))^n$ approaches $0$ exponentially with $n$ for $T
% \theta \ne p \pi$, which is usually the case, as $\theta$ is unknown.
% As a consequence, the SNR of the outcome is weakened, especially for
% $n \gg 1$. 

In this paper we propose a different method to perform multi-parameter
estimation that achieves the QML scaling $T$.  Interestingly, our
method focuses on the evolution of observables such as tensor products
of Pauli operators (i.e., we work in the Heisenberg picture), rather
than the evolution of the state of the probe itself.  For this reason,
in Sec.~\ref{lie} we start by giving a brief description of
Hamiltonian evolution in terms of Lie algebras.  We then present an
adaptive Bayesian estimation method to estimate single parameters at
the QML with DQC1 (Sec.~\ref{bayes}).  Moreover, in Sec.~\ref{mpe} we
show that, by applying dynamical-decoupling techniques and different
Suzuki-Trotter approximations, multi-parameter estimation can also be
performed with DQC1.  Here, we deduce that when the amount of
short-time evolutions is considered a resource, the QML is
asymptotically reached in the order of the approximation.  In
Sec.~\ref{rfa} we discuss the particular example of reference frame
alignment and show that to estimate the Euler angles
dynamical-decoupling techniques are not required. In Sec.~\ref{grover}
we discuss the reasons why DQC1 allows us to reach the fundamental
quantum limit in some cases, even though this model is less powerful
than standard quantum computation.  Finally, we present the
conclusions in Sec.~\ref{concl}.

In the following, we ignore the
effects of decoherence in our quantum algorithms and we assume that
all experimental parameters can be controlled with arbitrary
precision.

%%%%%%%%%%%%%%%%%%%%%%%%%%%%%%%%%%%%%
%%%%%%%%%%%%%%%%%%%%%%%%%%%%%%%%%%%%%

\section{Single-parameter estimation}
\label{spe}
When the $n$-qubit probe interaction can be described by a Hamiltonian
$H= \theta H_0$, single-parameter estimation aims to return an
estimate $\hat \theta$ of the unknown $\theta$ at the highest
precision possible, for some given amount of resources.  For fixed $n$
and $\parallel H_0\parallel$, our main resource is determined by the
total evolution time under $H$.  Let $W(T)=e^{-i\theta H_0 T}$ be the
(unitary) evolution operator induced by $H$, during a time interval
$T$. Clearly, if $W(T)$ acts non-trivially on some operator $O$,
information about $\theta$ can be gained by computing $h(\theta T)=\tr
[W^\dagger(T) O W(T) O]/2^n \in \mathbb{C}$ for different values of
$T$. The form of $h( \theta T)$ can be obtained through the
representation theory of Lie algebras (Sec.~\ref{lie}).  Contrary to
the example given in Sec.~\ref{intro}, $h( \theta T)$ can be estimated
using DQC1 without controlling the operation $W(T)$ (Sec~\ref{circuit}).  
Assuming that the accuracy in the estimation of
$h( \theta T)$ remains constant regardless of $T$, an adaptive
Bayesian estimator that returns $\theta$ at accuracy $\cO(1/T)$ (i.e.,
the QML) can be built in some cases of interest.  Furthermore, to
avoid signal loss due to possible large values of $n$, we choose the operator $O$ such
that $h( \theta T)$ does not depend on $n$.  These points are studied
and explained in more detail below.

%%%%%%%%%%%%%%%%%%%%%%%%%%%%%%%%%%%%%%%%%%%%%%%
%%%%%%%%%%%%%%%%%%%%%%%%%%%%%%%%%%%%%%%%%%%%%%%
\subsection{The Heisenberg picture}
\label{lie}
If $X=X^\dagger$ is an observable acting on $n$ qubits, we define
(Heisenberg picture)
\begin{equation}
\label{heis}
X(T)=W^\dagger(T) X W(T)\;,
\end{equation}
with $W(T)=e^{-i \theta H_0T}$ as above.  Assume now that we are given
two observables $H_1$ and $H_2$ such that, together with $H_0$, they
generate an $\fsu(2)$ Lie algebra. We obtain (see
appendix~\ref{ap:lie})
\begin{equation}
\label{su2}
H_1(T)= W^\dagger(T) H_1 W(T) = \cos (2 \theta T) H_1 - \sin (2 \theta T) H_2\;.
\end{equation}
If the operators are Schmidt pseudo-orthogonal 
($\tr[H_iH_j]=d \delta_{ij}$) we have
\begin{eqnarray}
  \cos(2 \theta T) &=& \tr[H_1(T) H_1]/d\;, \\
  \sin(2 \theta T) &=& -\tr[H_1(T) H_2]/d\;. \label{eq:sinest}
\end{eqnarray}

With no loss of generality we expand $H_j = \sum_{\mu=1}^L e^{\mu,j}
\sigma_{\mu,j}$, where $\sigma_{\mu,j} = \sigma_{\mu,j}^1 \otimes
\cdots \otimes \sigma_{\mu,j}^n$ are tensor products of Pauli
operators (henceforth simply called Pauli products), $\sigma_{\mu,j}^i
\in \{ \one, \sigma_x, \sigma_y, \sigma_z \}$, and $e^{\mu,j} \in
\mathbb{R}$ are known coefficients~\cite{comment0}.  We obtain
\begin{align}
\label{coslc}
\cos(2\theta T) &= \sum_{\mu,\mu'} e^{\mu,1} e^{\mu',1} \tr [ W^\dagger(T) \sigma_{\mu,1} W(T) \sigma_{\mu',1}]/d\;,\\ 
\label{sinlc}
\sin(2\theta T) &=- \sum_{\mu,\mu'} e^{\mu,1} e^{\mu',2} \tr [ W^\dagger(T) \sigma_{\mu,1} W(T) \sigma_{\mu',2}]/d\;.
\end{align}
That is, $\cos(2 \theta T)$ and $\sin(2 \theta T)$ can be estimated at
a fixed precision by $L^2$ runs of the circuit of Fig.~\ref{dqc1fig}.
Each run returns an estimate of $\tr[U_{\mu,j ; \mu',j'}]/2^n$, with
unitary $U_{\mu,j;\mu',j'} = W^\dagger(T) \sigma_{\mu,j} W(T)
\sigma_{\mu',j'}$. In Sec.~\ref{bayes} we show how to estimate
$\theta$ at the QML from the estimation of $\cos(2 \theta T)$ or
$\sin(2 \theta T)$, for different values of $T$. 
Although we are mainly interested in the fundamental
scaling achieved by increasing $T$,
it is worth noting that 
the scaling with $L^2$  can be
largely reduced in some cases of interest (see
Appendix~\ref{ap:lie}).

%%%%%%%%%%%%%%%%%%%%%%%%%%%%%%%%%%%%%%%%%%%%%%%%
%%%%%%%%%%%%%%%%%%%%%%%%%%%%%%%%%%%%%%%%%%%%%%%%
\subsection{DQC1 circuits}
\label{circuit}
\begin{figure}[!h]
\begin{equation*}
  \Qcircuit @C=.5em @R=-.5em {
      \lstick{\ketbra{0} {\sf a}}  & \gate{H} &\ctrl{1}  & \ctrl{1}  & \ctrl{1}  & \ctrl{1} & \meter  & \qw & \push{\rule{0em}{4em}} \\
   & \qw & \multigate{4}{\sigma_{\mu',j'}}  &    \multigate{4}{W(T)}  &    \multigate{4}{\sigma_{\mu,j}}  &    \multigate{4}{W^\dagger(T)} & \qw &  \\
    &  \qw &  \ghost{\sigma_{\mu',j'}}  &    \ghost{W(T)}  &    \ghost{\sigma_{\mu,j}}  &    \ghost{W^\dagger(T)} & \qw & \\
    \lstick{\mbox{$\one_{n}/2^{n}$}}  & \qw &   \ghost{\sigma_{\mu',j'}}  &    \ghost{W(T)}   &   \ghost{\sigma_{\mu,j}}  &    \ghost{W^\dagger(T)} & \qw & \\
    &   \qw & \ghost{\sigma_{\mu',j'}}  &    \ghost{W(T)}  &    \ghost{\sigma_{\mu,j}}  &    \ghost{W^\dagger(T)} & \qw & \\
    &  \qw &  \ghost{\sigma_{\mu',j'}}  &    \ghost{W(T)}  &    \ghost{\sigma_{\mu,j}}  &    \ghost{W^\dagger(T)} & \qw   \gategroup{2}{2}{6}{1}{.5em}{\{} 
  }
\end{equation*}
\caption{DQC1 circuit for the estimation of $\tr[W^\dagger(T) \sigma_{\mu,j} W(T) \sigma_{\mu',j'}]/2^n
\equiv \langle \sigma_x^{\sf a} \rangle$. Note that $\langle  \sigma_y^{\sf a}\rangle=0$ in this case. Since every unitary is controlled
by ${\sf a}$ (filled circles), including the evolution operator $W(T)$, the execution of this algorithm may be unfeasible due to the nature of the problem.}
\label{dqc1fig2}
\end{figure}
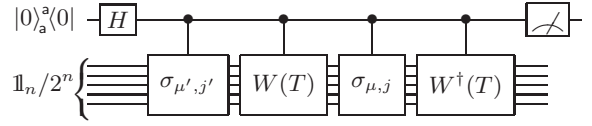

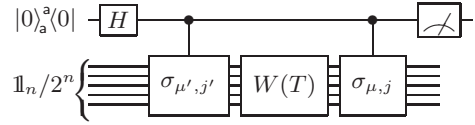
\begin{figure}[!h]
\begin{equation*}
  \Qcircuit @C=.5em @R=-.5em {
      \lstick{\ketbra{0}{\sf a}}  & \gate{H}  &\ctrl{1}  & \qw  & \ctrl{1} & \meter  & \qw & \push{\rule{0em}{4em}} \\
   &  \qw &  \multigate{4}{\sigma_{\mu',j'}}  &    \multigate{4}{W(T)}  &    \multigate{4}{\sigma_{\mu,j}}   & \qw &  \\
    &   \qw &   \ghost{\sigma_{\mu',j'}}  &    \ghost{W(T)}  &    \ghost{\sigma_{\mu,j}}  &    \qw & \\
    \lstick{\mbox{$\one_{n}/2^{n}$}}  &  \qw&    \ghost{\sigma_{\mu',j' }}  &    \ghost{W(T)}  &    \ghost{\sigma_{\mu,j}}& \qw & \\
    &   \qw &   \ghost{\sigma_{\mu',j'}}  &    \ghost{W(T)}  &    \ghost{\sigma_{\mu,j}}  &  \qw & \\
    &   \qw &   \ghost{\sigma_{\mu',j'}}  &    \ghost{W(T)}  &    \ghost{\sigma_{\mu,j}}  &   \qw   \gategroup{2}{2}{6}{1}{.5em}{\{} 
  }
\end{equation*}
\caption{Simplified version of the DQC1 circuit of Fig.~\ref{dqc1fig2}. The operators $\controlu W(T)$ and $W^\dagger(T)$ are avoided. The estimation of $\cos(2 \theta T)$ or $\sin(2 \theta T)$ [Eqs.~(\ref{coslc}) and~(\ref{sinlc})] is performed using available resources only.}
\label{dqc1fig3}
\end{figure}

We now show how to avoid the need of controlled $\controlu W(T)$ operations, 
when estimating parameters with DQC1. This
is of great importance since $\controlu W(T)$ may not be available for
our use.  To show this, we focus on the estimation of
$\tr[W^\dagger(T)\sigma_{\mu,j} W(T) \sigma_{\mu',j'}]$, as required
by Eqs.~(\ref{coslc}) and~(\ref{sinlc}).  The circuit that
accomplishes this task is shown in Fig.~\ref{dqc1fig2}. Here, the
$\controlu U$ operation of Fig.~\ref{dqc1fig} has been replaced by
$\controlu W^\dagger(T) \controlu \sigma_{\mu,j} ~ \controlu W(T)
\controlu \sigma_{\mu',j'}$, where each operation is controlled by the
ancilla ${\sf a}$. Nevertheless, note that one can accomplish the same
task even if the action of the operators $W(T)$ and $W^\dagger(T)$ is
not controlled~\cite{SOG02}.  That is,
\begin{equation}
  \controlu W^\dagger(T) \controlu \sigma_{\mu,j} \, \controlu W(T) \controlu \sigma_{\mu',j'} \equiv W^\dagger(T) \controlu \sigma_{\mu,j} W(T) \controlu \sigma_{\mu',j'} \;,
\end{equation}
which is clearly the identity operator when $\sf a$ is in
$\ket{0}_{\sf a}$. In Fig.~\ref{dqc1fig3} we show a simplified circuit
that allows us to compute the above trace.  The last operation
$W^\dagger(T)$ is not included as it does not alter the measurement
outcome. Also, $W^\dagger(T)$ may not be an available resource.  Thus,
the circuit can be implemented using (known) elementary gates and the
available time-evolution operator $W(T)$ only.

The next step is to detail a strategy to estimate $\theta$ from
$\cos(2 \theta T)$ or $\sin(2 \theta T)$, and to characterize the associated error and resources
needed. For this reason, we first make an assumption on 
the standard deviation $\Delta$ of the output  returned by the DQC1 circuit of Fig.~\ref{dqc1fig3}.
Obviously, the smaller $\Delta$ is, the better the precision of the
resulting estimation.  We consider $\Delta \ll 1$, which can always be
achieved by simple repetition of the computation.  In some cases, like
liquid-state NMR quantum computation, where a vast amount of molecules
contribute to the output signal, a $\Delta \ll 1$ could be achieved in a
single run.  Our estimation procedure should take advantage of this
property by going beyond just performing a bit by bit estimation of
$\theta$, as it is done in several pure-state phase estimation
techniques that involve strong (projective)
measurements~\cite{KOS07,VARIOUS1}.  Moreover, we assume that $\Delta$
remains approximately constant in a certain region of values of $2\theta
T$.
This is a consequence of
weak measurements~\cite{LS00}.   In view of the central limit theorem, we assume 
that the measurement outcome is distributed according to $\cN(\cos(2 \theta T), \Delta^2)$ or
$\cN(\sin(2 \theta T), \Delta^2)$
(see Sec.~\ref{intro} and Ref.~\cite{comment-1}).

%%%%%%%%%%%%%%%%%%%%%%%%%%%%%%%%%%%%%%
%%%%%%%%%%%%%%%%%%%%%%%%%%%%%%%%%%%%%%
\subsection{Adaptive Bayesian estimation}
\label{bayes}
In Bayesian estimation, a parameter $\alpha$ to be estimated
is considered to be a
variable with an associated (known) probability distribution $f(\alpha)$
(i.e., the {\em prior distribution}).  The prior distribution
formalizes the experimenter's \emph{state of belief} about $\alpha$. It
is the job of the experimenter to gain access to a sample of data
$\{x_1,\dots,x_K\}$ whose distribution $f(x_K,\ldots,x_1 | \alpha)$ depends on
$\alpha$. Thus, the joint distribution of $\{x_1,\dots,x_K\}$ and
$\alpha$ is
\begin{align}\label{eq:ftn}
  f( x_K,\ldots,x_1,\alpha) = f(x_K,\ldots,x_1 | \alpha) f(\alpha)\;.
\end{align}
After observing a set of measurement outcomes $\{x_1,\ldots,x_K\}$, this
information is used to obtain a
\emph{posterior distribution} $f(\alpha | x_K,\ldots,x_1)$ that
corresponds to the experimenter's updated state of belief about the
unknown $\alpha$. This update is done using
Bayes' rule
\begin{align}
\label{eq:fna}
  f(\alpha | x_K, \ldots, x_1 ) = \frac { f( x_K,\ldots,x_1,\alpha)  } { f(x_K, \ldots, x_1)} \;,
\end{align}
where \emph{the marginal sampling} distribution $ f(x_K, \ldots, x_1)$ can
be calculated from the joint distribution as
\begin{align}\label{eq:fnx}
  f(x_K, \ldots, x_1) = \int f(x_K, \ldots, x_1, \alpha) d\alpha\;.
\end{align}

In standard Bayesian estimation the probability
of observing an {\em i.i.d.} sample 
$\{x_1,\dots,x_K\}$ is determined by the total sampling distribution
$f(x_K, \ldots, x_1|\alpha) =f(x_K |
\alpha) \cdots f(x_1 | \alpha) $. In \emph{adaptive Bayesian
  estimation} the outcome of the first measurement $x_1$ can be used
to control the sampling distribution of the second measurement,
$f(x_2|x_1, \alpha)$ (see below). 
In general, the sampling distribution of the $l$th measurement outcome
can be conditioned by $\{x_1,\ldots,x_{l-1}\}$. At the end,
$f(x_K,\ldots,x_1| \alpha)= f(x_K | x_{K-1},\ldots,x_1, \alpha) \cdots f(x_1|\alpha)$.

The Bayes' risk quantifies the expected penalty to be paid when
 using a particular estimator $\hat
\alpha(x_K,\ldots,x_1)$ of $\alpha$, and a given {\em cost function}. It is common to
search for an $\hat \alpha$ that minimizes the Bayes'
risk with a quadratic cost function,  given by
\begin{align}
  \int (\alpha - \hat \alpha(x_K, \ldots, x_1) )^2 f(\alpha | x_K,
  \ldots, x_1 ) d\alpha\;.
\end{align}
This risk is just the variance of the estimator and is
minimized by the expectation value of the posterior distribution,
giving the optimal estimator
\begin{align}\label{eq:ha}
  \hat \alpha(x_K,\ldots,x_1) = \int \alpha f(\alpha | x_K, \ldots,
  x_1 ) d \alpha\;.
\end{align}

While a more detailed explanation of the adaptive Bayesian procedure is
given in Appendix~\ref{ap:bayes}, in the following we present a generic step
of the estimation. 
Denote by $\Delta \ll 1$  the output precision
of DQC1 when measuring $\cos(2 \theta T)$, which may actually involve
many ($L^2>1$) different runs of the circuit of Fig.~\ref{dqc1fig3}.
Assume that, from $l$ previous estimations, we have obtained an
estimator $\hat \theta_{l}$ of $\theta$ such that, for known evolution
time $T_{l}$
and integer $p_{l}$,
\begin{align}
\label{thetal}
  2 \hat \theta_{l} T_{l} \approx \pi/2 + 2 p_{l} \pi\;.
\end{align}
Furthermore, assume that the 95\% confidence interval for the estimation is
\begin{align}
\label{deltal}
  \hat \theta_{l} - 1.96 \Delta_{l}/(2 T_{l}) \le \theta \le \hat \theta_{l} + 1.96 \Delta_{l}/(2 T_{l})\;,
\end{align}
with $\Delta_{l} < \Delta$. Next, we show how to
{\em zoom in} on $\hat \theta_l$ to
obtain an estimator $\hat \theta_{l+1}$, so  
Eqs.~(\ref{thetal}) and~(\ref{deltal}) are still satisfied when replacing $l \rightarrow l+1$.

To do this,  we first find $T_{l+1}$ such that
\begin{equation}
2 \hat \theta_{l} T_{l+1} =  \pi/2 + 2p_{l+1} \pi\;,
\end{equation}
with $p_{l+1} > p_{l}$ integer. 
The $(l+1)$th measurement returns $x_{l+1}$, an estimate
of $\cos(\alpha_{l+1})$, with $\alpha_{l+1} = 2 \theta T_{l+1}$. This
is done by running the algorithm of Fig.~\ref{dqc1fig3} with
$T=T_{l+1}$. We chose $p_{l+1}$ close enough to $p_{l}$ such that, for
$\alpha'_{l+1} = \alpha_{l+1} - (\pi/2 + 2 p_{l+1} \pi)$, we
approximate 
$  \cos{\alpha_{l+1}} \approx -\alpha'_{l+1}$
(see Appendix~\ref{ap:bayes}).
Using Bayes rule and the joint distribution for the $(l+1)$th measurement, we obtain
\begin{align}
\label{dens8}
f( x_{l+1}, &\alpha_{l+1}| x_{l},\ldots, x_1) \\
&= f(x_{l+1}|\alpha_{l+1},x_l,\cdots,x_1)
f(\alpha_{l+1}| x_l,\cdots,x_1) \nonumber \\
\nonumber
&\approx
\frac 1 { \sqrt{2 \pi} \Delta } e^{-\frac {(x_{l+1} + \alpha'_{l+1} )^2}{2
    \Delta^2}} \frac 1 { \sqrt{2 \pi} a_{l+1} \Delta_{l} } e^{-\frac
  {(\alpha'_{l+1} )^2}{2 (a_{l+1})^2 \Delta_{l}^2}}\;,
\end{align}
with $a_{l+1} = T_{l+1}/T_{l}$.  Equation~(\ref{dens8}) determines the posterior distribution 
$f(\alpha_{l+1}
| x_{l+1},x_{l},\ldots,x_1)=f( x_{l+1}, \alpha_{l+1}| x_{l},\ldots, x_1) /f(x_{l+1}|x_l,\cdots,x_1)$. This adaptive procedure returns a new
estimator $\hat \theta_{l+1}$, and standard deviation $\Delta_{l+1} / (2T_{l+1})$, with $\Delta_{l+1}<\Delta$,
satisfying
\begin{equation}
\label{l+1est}
 \hat \theta_{l+1} -1.96 \Delta_{l+1}/(2T_{l+1}) \le \theta \le \hat \theta_{l+1} + 1.96 \Delta_{l+1}/(2T_{l+1})\;.
\end{equation}

The total number $K$ of estimations is chosen such that the final
standard deviation is reduced below the desired precision. A
sufficient condition is $\Delta / (2T_K) \le \Delta_\theta$.  The fact
that the standard deviation is reduced by $T_l$ at each step [Eq.~(\ref{l+1est})] 
  guarantees that the QML is
achieved (Appendix~\ref{ap:bayes}).  

Remarkably, the confidence level of estimating the mean with error
$\epsilon$ in our algorithm increases exponentially as
$1-e^{-C(\epsilon/\tau)^2)}$, with $C>0$ and $\tau$ the corresponding
standard deviation.  This is clearly an advantage with respect to the
standard pure-state phase estimation algorithm, where the confidence
increases as $1-\cO(1/\epsilon)$~\cite{CEMM98}.

%%%%%%%%%%%%%%%%%%%%%%%%%%%%%%%%%%%%%%%%%%
%%%%%%%%%%%%%%%%%%%%%%%%%%%%%%%%%%%%%%%%%%
\subsection{Black-box estimation: discrete time evolution}
\label{blackbox}
Imagine now that, instead of being able to evolve under the action
of $H$ for any period of time $T$, we are given a black box whose
action is to perform the unitary operation $W_{B}=e^{-iH}$. Like in
the previous case,
$H=\theta H_0$. That is, we are only allowed to evolve under $H$ for a
discrete time by simple concatenation of $W_{B}$'s. This condition restricts
the set of accessible operations to elementary gates and operations of the form
\begin{equation}
\bar{W}(q) =  \overbrace{ W_{B} \cdots W_{B}}^{\mbox{$q$ times}} \;,
\end{equation}
only. We seek to estimate $\theta$ at the QML
using a modification of the previous adaptive Bayesian method.

We now give the generic step for achieving the QML in the discrete
time case (see Appendix~\ref{ap:discrete} for the first step).  We
assume that $\hat \theta_l$ is the mean of the estimator obtained
after the $l$th measurement, performed with $q_l \in \mathbb{N}^*$ uses
of $W_B$.  Because we seek to make estimations around $\pi/2$, we take
\begin{equation}
2 \hat\theta_{l-1} q_l + 2\phi_l = \pi/2 + 2p_l \pi, \ \forall l \;,
\end{equation}
with $\hat \theta_{l-1}$ the mean of the estimation in the $(l-1)$th
measurement, $p_l \in \mathbb{N}^*$, and $\phi_l$ the phase
compensation. In other words, the $l$th estimation was performed by
implementing the algorithm of Fig.~\ref{dqc1fig3} with $W(T)
\rightarrow W_a(q_l,\phi_l) \equiv (W_B)^{q_l} e^{-i \phi_l H_0}$.
Because we can always consider $\pi/2 \ge \phi_l > -\pi/2$, the phase
compensation is made with unit cost.  We also assume that $\Sigma_l
\le \Delta$ is the standard deviation of the estimator of $2 \theta
q_l$~\cite{comment8}.

For the $(l+1)$th measurement, we write $q_{l+1} = b q_l$ with
small enough $b$~\cite{comment9}. The $(l+1)$th measurement
returns $y_{l+1}$, an estimate of $\cos(\beta_{l+1})$, with
$\beta_{l+1} = 2 \theta q_{l+1} + 2 \phi_{l+1} \approx \pi/2 +2 \pi
p_{l+1}$. Thus, to obtain $f(\beta_{l+1}|y_{l+1},\ldots,y_1)$ in the
adaptive Bayesian step, we approximate
\begin{align}
\label{bbdens2}
f(y_{l+1}|y_l, \ldots, y_1, \beta_{l+1}) \approx \frac 1 { \sqrt{2
    \pi} \Delta } e^{-\frac {(y_{l+1} + \beta'_{l+1} )^2}{2
    \Delta^2}}\;,
 \end{align}
with 
\begin{equation}
\beta'_{l+1} = \beta_{l+1} - (\pi/2 + 2p_{l+1} \pi)\;.
\end{equation}
Moreover, since
\begin{equation}
\label{bbdens3}
f(\beta_{l+1}|y_{l}, \ldots, y_1) = \frac 1 { \sqrt{2 \pi}  b\Sigma_l } e^{-\frac { (\beta'_{l+1}
)^2}{2 b^2 (\Sigma_l)^2}}\;,
\end{equation}
the resulting distribution $f(\beta_{l+1}|y_{l+1},\ldots,y_1)$ is normal. This is a consequence of Bayes' rule. 
Its mean and standard deviation, obtained by combining the exponents appearing in Eqs.~\eqref{bbdens2} and~\eqref{bbdens3}, are
\begin{eqnarray}
\label{bbdens4}
\hat{\beta}'_{l+1} &=&- \frac{(b'_{l+1})^2} {1+(b'_{l+1})^2 } y_{l+1} \;, \\
\label{bbdens5}
\Sigma_{l+1}&=&\left ( b'_{l+1}/\sqrt{1+(b'_{l+1})^2} \right)\Delta < \Delta\;, 
\end{eqnarray}
with $b'_{l+1} = b \Sigma_l/\Delta < b$.  Equations~\eqref{bbdens4} and~\eqref{bbdens5} guarantee the success of the induction method. These quantities determine the mean and standard deviation of the new (improved) estimator of $\theta$  as
\begin{eqnarray}
\nonumber
\label{bbdens6}
\hat \theta_{l+1}& =& \frac{1}{2b^l} \left( \pi/2 + 2p_{l+1} \pi -2 \phi_{l+1} -  \frac{(b'_{l+1})^2} {1+(b'_{l+1})^2 } y_{l+1} \right) \ ,  \\
\label{bbbdens7}
\Sigma'_{l+1} &=& \frac{ \Sigma_{l+1} }{2b^l} < \frac{\Delta}{2b^l}\;.
\end{eqnarray}
The similarity of these results and those obtained for the continuous time case [Eqs. ~\eqref{dens7a} and~\eqref{dens7}] is clear.

Summarizing, if $\Delta_\theta$ denotes the desired precision in the parameter estimation, 
a sufficient condition is to choose the total number of measurements $K$ such that the final precision
satisfies
$\Delta/(2b^{K-1})\le \Delta_\theta$. Since the total number of uses of $W_B$ is given by 
$1+\cdots + b^{K-1} = (b^K-1)/(b-1)=\cO[\Delta / (2\Delta_\theta)]$,  the QML is also reached in this case.

%%%%%%%%%%%%%%%%%%%%%%%%%%%%%%%%%%%%%
%%%%%%%%%%%%%%%%%%%%%%%%%%%%%%%%%%%%%

\section{Multi-parameter Estimation}
\label{mpe}
In this section we consider a more general case where the unknown interaction 
with  the $n$-qubit probe can be described by a Hamiltonian 
\begin{equation}
H= \sum_{\nu=1}^P  \theta^\nu \sigma_\nu\;.
\end{equation}
Here, $\theta^\nu \in \mathbb{R}$ 
and $\sigma_\nu$ are Pauli products. 
Using the results of Sec.~\ref{bayes}, we seek  to estimate every (unknown)  $\theta^\nu$ such that
the returned precision approaches the QML for a given amount of
resources or evolution time.
For this reason,  we assume that a previous estimate of every
parameter, with mean $\pi > \hat \theta_0 ^\nu >0$, is known.

Using  dynamical-decoupling techniques~\cite{VK05},
the multi-parameter estimation case can be converted into $P$ single-parameter estimations.  
For simplicity, we
consider first the case where there is a
Pauli product $\sigma$ such that (for some $\nu$)
\begin{eqnarray}
\left[ \sigma_\nu, \sigma \right] &=& 0 \; , \\
\left\{ \sigma_{\nu'} , \sigma \right\} &=& 0 \ \forall \  \nu' \ne \nu\;.
\end{eqnarray}
Then,
\begin{align}
  H_\nu= \theta^\nu \sigma_\nu = (H + \sigma H \sigma)/2\;.
\end{align}
(In general, methods like the one just described
can be used to decouple any $H_\nu$ from any $H$.)
We define $S_\nu(T) = e^{-i H_\nu T}$ to be the corresponding
evolution operator.  If such an operator were to be an available
resource, $\theta^\nu$ could be estimated using the scheme
of Sec.~\ref{bayes}. To do this, we would have to replace $W(T) \rightarrow S_\nu(T)$ and $\sigma_{\mu,j} \rightarrow \sigma_1$ in the
circuit of Fig.~\ref{dqc1fig3}, with $\{\sigma_\nu,\sigma_1\}$ being Pauli products that generate an $\fsu(2)$ algebra.

We now show how to approximate $S_\nu(T)$ from accessible operations
that include $W(T)$ and elementary gates only.  To show this, we use a
Suzuki-Trotter approximation~\cite{Suz85}.  Specifically, for
$q=1/\epsilon \in \mathbb{N}^*$, we decompose
\[
S_\nu(T)= \overbrace{ S_\nu(\epsilon T) \cdots S_\nu(\epsilon T)}^{\mbox{$q$ times}} \ .
\]
If $\bar{S}_\nu(\epsilon T)$ denotes a $p$th order Suzuki-Trotter
approximation to $S_\nu(\epsilon T)$, we have
\begin{equation}
\label{trot1}
\parallel S_\nu(\epsilon T)- \bar{S}_\nu(\epsilon T) \parallel = \cO[ \parallel H \parallel^p (\epsilon T)^p]\; ,
\end{equation}
with $\parallel . \parallel$ some operator norm (e.g., the largest eigenvalue).
Then,
\begin{equation}
\label{trot2}
\varepsilon/2 \equiv \parallel S_\nu (T) - \bar{S}_\nu(T) \parallel = \cO[ \parallel H \parallel ^p\epsilon^{p-1} T^p]\; ,
\end{equation}
with $\bar{S}_\nu (T) = [\bar S_\nu(\epsilon T)]^q$.
Equation~\eqref{trot2} was obtained using Eq.~\eqref{trot1}, together with
$\parallel A^q- B^q \parallel= \parallel (A-B) A^{q-1} + B (A-B)
A^{q-2} + \ldots + B^{q-1} (A-B)\parallel \le q \parallel A-B \parallel$, for $A$ and $B$ unitaries.

We now show how to build $\bar S_\nu (T)$ out of available resources. In the simplest case
(i.e., $p=2$) the evolution operator at short times factorizes as
\begin{equation}
\label{trot3}
 \bar{S}_{\nu}(\epsilon T)= e^{-i H \epsilon T/2} \sigma e^{-i  H \epsilon T/2} \sigma\;.
 \end{equation}
Then, $\bar{S}_\nu(T) $ can be implemented using $H$
evolutions and $\sigma$ gates only.   Similarly, a second-order Suzuki-Trotter approximation
(i.e., $p=3$) is given by
\begin{equation}
  \bar{S}_{\nu}(\epsilon T)= e^{-i H \epsilon T/4} \sigma e^{-i H \epsilon T/2} \sigma e^{-i H \epsilon T/4}\;.
\end{equation}
Higher order approximations can be constructed in a similar fashion~\cite{Suz92},
so they can always be implemented with accessible gates. 
The larger $p$ is, the shorter the actions of $H$ in each step. Thus, 
we require precise time control in our algorithms (measurements).

Replacing $W(T) \rightarrow \bar S_\nu(T)$ and $\sigma_{\mu,j}
\rightarrow \sigma_1$ in the circuit of Fig.~\ref{dqc1fig3} allows us
to estimate $\theta^\nu$. To show this, consider the measurement
output $z_l$ obtained in the $l$th measurement.  $z_l$ will give us an
estimator of the angle $\theta^\nu$ plus a correction
\begin{equation}
 \cos(2 \theta^\nu T_l) + \gamma(\epsilon,p,T_l,\overrightarrow \theta)\;,
\end{equation}
with $\overrightarrow \theta= \theta^1,\ldots,\theta^P$.
The norm of  $\gamma(\epsilon,p,T_l,\overrightarrow \theta) \in \mathbb{R}$ can be bounded above as
\begin{align}
\nonumber
&\ \ \ \ \ \  |\gamma(\epsilon,p,T_l,\overrightarrow \theta) | = \\
\nonumber 
&= \frac{ |\tr [S^\dagger_\nu(T_l)  \sigma_1 S_\nu(T_l) \sigma_1   
-\bar S^\dagger_\nu(T_l) \sigma_1  \bar S_\nu(T_l) \sigma_1]|}{2^n}\\
\label{gammanorm}
&\le\frac{ \tr \parallel S^\dagger_\nu(T_l)  \sigma_1 S_\nu(T_l) \sigma_1 - \bar S^\dagger_\nu(T_l) \sigma_1  \bar S_\nu(T_l) \sigma_1\parallel}{2^n} \le \varepsilon\;.
\end{align}
We have used Eq.~\eqref{trot2} to obtain Eq.~\eqref{gammanorm}. 
Because $\gamma(\epsilon,p,T_l,\overrightarrow \theta)$ depends on the $\theta^{\nu}$'s, we consider it 
a variable with an associated (worst-case scenario) prior distribution given by $f(\gamma) \equiv
\cN(0,\Delta_\gamma)$, with $\Delta_\gamma =\cO( \Delta \varepsilon)$~\cite{comment7}.
The net effect in the adaptive Bayesian procedure is that  now,  the joint distribution
determines a marginal distribution after $\gamma(\epsilon,p,T_l,\overrightarrow \theta)$ is integrated out. More precisely, for the $l$th estimation, we have
\begin{align}
\nonumber
  & f(x_l | x_{l-1},\ldots,x_1, \alpha^\nu_l)=  \\
\label{intgamma1}
= & \int f(x_l |
  x_{l-1},\ldots,x_1, \alpha^\nu_l,\gamma) f(\gamma) d\gamma \;,
 \end{align} 
 with $\alpha^\nu_l = 2 \theta^\nu T_l$.
Making a linear approximation in the cosine function, we obtain
\begin{equation}  
  \label{intgamma2}
f(x_l | x_{l-1},\ldots,x_1, \alpha^\nu_l)\approx  \cN(\alpha'^\nu_l, (\Delta')^2)\;,
\end{equation}
with  $\alpha'^\nu_l = \alpha^\nu_l - (\pi/2 +2p_l \pi)$ (see Sec.~\ref{bayes}), and updated variance 
\begin{equation}
(  \Delta')^2 = \Delta^2 + \Delta_\gamma^2 \; .
\end{equation}
Thus, we can use the adaptive method of Sec.~\ref{bayes} to estimate every parameter $\theta^\nu$ at the QML, by replacing $\Delta \rightarrow \Delta'$.

Notice that one could also implement an adaptive Bayesian approach to learn about $\gamma(\epsilon,p,T_l,\overrightarrow \theta)$. In such a case, the distribution $f(\gamma)$ would need to be updated after each measurement based on the previous measurement outcomes. Nevertheless, we did not consider this approach in the above discussion because we assumed that the Suzuki-Trotter approximation used is good enough for our purposes (i.e., $ \Delta_\gamma \ll 1$).

When the resource of interest is  the number of calls to $\bar S_\nu (\epsilon T)$,
the amount of resources to reach a precision $\Delta_{\theta^\nu}$ changes.
For this reason, consider the total evolution time $T_t = \cO[(\Delta + \Delta_\gamma)/\Delta_{\theta^\nu})]$, with $\Delta_\gamma = \cO(\Delta \parallel H \parallel^p \epsilon^{p-1} T_t^p)$.
Assume that we want to keep $T_t$ constant, regardless of $p$. Then, $q=1/\epsilon =\cO[(T_t)^{(p/p-1)}]$.
That is, $\cO[1/(\Delta_{\theta^\nu})^{(p/p-1)}]$ actions of $\bar S_\nu(\epsilon T)$ are required to attain  $\Delta_{\theta^\nu}$, and the  QML is asymptotically reached in $p$.

%%%%%%%%%%%%%%%%%%%%%%%%%%%%%%%%%%%%%%%%%%%%%%%
%%%%%%%%%%%%%%%%%%%%%%%%%%%%%%%%%%%%%%%%%%%%%%% 

\section{Reference frame alignment}
\label{rfa}
Imagine that two distant parties, Alice and Bob, suffer some frame
misalignment, which is manifest in the way they characterize their
operations on equivalent quantum systems.  This might be a result of
not sharing synchronized clocks or having different spatial reference
frames.  Aligning both frames requires the exchange of physical
systems carrying ``unspeakable'' information. This information is
encoded as frame-dependent parameters that need to be
estimated~\cite{frames}.  The resource that limits the quality of this
estimation is the number of systems interchanged between Alice and
Bob.  We will show that Alice and Bob can align their frames within
the DQC1 model at the QML.  In particular, we propose a modification of a
pure-state protocol for frame synchronization~~\cite{VARIOUS1} based
on repeated coherent exchanges of the $n$-qubit probe only.
Remarkably, the state of the probe remains completely mixed and
separable from the ancilla at every step~\cite{VARIOUS2}. Moreover, in
our protocol, Bob never accesses the ancilla that Alice measures.

We consider first the case where the effect of the frame
misalignment is uni-parametric. That is, Alice's and Bob's description
of operators acting on equivalent  Hilbert spaces is known to
differ by a unitary transformation $V_\theta = e^{-i \theta H_0}$, with
$\theta$ unknown. More explicitly, for some operator $O$ we have
\begin{align}
  O^B = V^\dagger_\theta O V_\theta = e^{i \theta H_0}
  O e^{-i \theta H_0}\;,
\end{align}
where we used the superscript $B$ to denote the action of $O$ in Bob's
frame. Since $H_0$ is known, we assume that we can find
pseudo-orthogonal observables $H_1$ and $H_2$ such that they form an
$\fsu(2)$ algebra [i.e., Eq.~\eqref{eq:2ie} is satisfied in Alice's
frame].

\begin{figure}[!h]
\begin{equation*}
  \Qcircuit @C=.5em @R=-.5em {
      \lstick{\ketbra{0}{\sf a}}  & \gate{H}  &\ctrl{1}  & \qw  \raisebox{-.9cm}{\mbox{Bob}} & \qw  & \ctrl{1} & \meter  & \qw & \push{\rule{0em}{5em}} \\
   &  \qw &  \multigate{4}{\sigma_{\mu',j'}}  &    \multigate{4}{e^{-i \frac{\pi}{2} H_1^B }} &    \multigate{4}{e^{i \frac{\pi}{2} H_1 }}   &    \multigate{4}{\sigma_{\mu,j}}   & \qw &  \\
    &   \qw &   \ghost{\sigma_{\mu',j'}} &    \ghost{e^{-i \frac{\pi}{2} H_1^B }} &    \ghost{e^{i \frac{\pi}{2} H_1 }}   &    \ghost{\sigma_{\mu,j}}  &    \qw & \\
    \lstick{\mbox{$\one_{n}/2^{n}$}}  &  \qw&    \ghost{\sigma_{\mu',j' }}  &    \ghost{e^{-i \frac{\pi}{2} H_1^B}} &    \ghost{e^{i \frac{\pi}{2} H_1 }}  &    \ghost{\sigma_{\mu,j}}& \qw & \\
    &   \qw &   \ghost{\sigma_{\mu',j'}}  &   \ghost{e^{-i \frac{\pi}{2} H_1^B }} &   \ghost{e^{i \frac{\pi}{2} H_1 }} &      \ghost{\sigma_{\mu,j}}  &  \qw & \\
    &   \qw &   \ghost{\sigma_{\mu',j'}}  &     \ghost{e^{-i \frac{\pi}{2} H_1^B }}  &  \ghost{e^{i \frac{\pi}{2} H_1 }} &     \ghost{\sigma_{\mu,j}}  &   \qw   \gategroup{2}{2}{6}{1}{.5em}{\{} \gategroup{2}{4}{6}{4}{.5em}{--}
  }
\end{equation*}
\caption{Elementary step of the frame alignment protocol. 
The $n$-qubit probe, whose state remains completely mixed,
 is exchanged between Alice and Bob. The Pauli products $\sigma_{\mu,j}$
 and $\sigma_{\mu',j'}$ appear in the decomposition of $H_1$ 
 [see Eqs~\eqref{coslc} and \eqref{sinlc}].}
\label{fselem}
\end{figure}
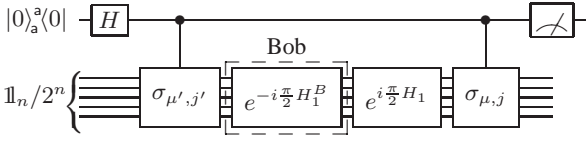

An elementary step of the protocol starts with Alice sending the $n$-qubit probe
to Bob.  Subsequently, Bob applies the operation $e^{-i \pi H_1^B /2}$
and returns the probe to
Alice. Finally, Alice applies the adjoint operation $e^{i \pi H_1
  /2}$. The resulting operation on the state of the probe in Alice's frame is
\begin{align}
\nonumber
  e^{i \pi H_1 /2} e^{-i \pi H_1^B /2} &=  e^{i \pi H_1 /2} e^{i \theta H_0} e^{-i \pi
    H_1 /2} e^{-i \theta H_0} \\ 
     \label{eq:lieresult}
    &= e^{-2 i \theta H_0} \equiv
  V_{2 \theta}\;.
\end{align}
Equation~\eqref{eq:lieresult} can be obtained by working in any faithful
representation of $\fsu(2)$. The global action of each step 
can be seen as a ``black box'' implementation of 
$V_{2 \theta}$, whose parameter we want to estimate.
Using the results of Sec.~\ref{blackbox}, the circuit of Fig.~\ref{dqc1fig3} can be used to make
a first estimate of $\theta$ if $W(T)$ is replaced by $V_{2\theta}$ (see Fig.~\ref{fselem}).
To zoom in on previous estimations requires instead the implementation of the unitary
$V_{2 m \theta}=(V_{2\theta})^m$, with $m \in \mathbb{N}^*$. This can be done by simple concatenation
of elementary steps, requiring $m$ coherent exchanges of the probe.

Another case of interest is the alignment of  spatial
reference frames~\cite{VARIOUS1}. Let us assume that Bob's operators
are related to Alice's through a rotation  
\begin{align}
  R = e^{ - i \psi H_2} e^{-i \theta H_1} e^{-i \phi H_2}\;,
\end{align}
where $\{\phi,\theta, \psi\}$ are Euler angles and $\{ H_1, H_2\}$
generate an $\fsu(2)$ algebra. Consider a synchronization
protocol with this elementary step: Alice sends the probe to Bob; Bob
applies the operation $e^{-i \pi H^B_2 /2}$ and returns the probe to
Alice; Alice applies the operation $e^{i \pi H_2 /2}$. 
The effective rotation of this step, in Alice's frame, is given by
\begin{align}
  V' = e^{i \pi H_2 /2} e^{-i \pi H^B_2 /2} = 
  e^{i \pi H_2 /2} R^\dagger e^{-i \pi H_2 /2} R
  \;.
\end{align}
The action of $V'$ on $H_2$ implies that
\begin{align}
  \tr[ V'^\dagger H_2 V' H_2] = d \cos(2 \theta)\;,
\end{align}
with $d$ some normalization constant [Eq.~\eqref{coslc}].
Again, the previous derivation can be carried out in any faithful representation of $\fsu(2)$.
Thus, a first estimation of the Euler angle $\theta$ can be performed by using the circuit of Fig.~\ref{dqc1fig3}, replacing $W(T) \rightarrow V'$, and with $\sigma_{\mu,j}$ being the Pauli products appearing in the expansion of $H_2$.
Furthermore, since
\begin{align}
  \tr[ (V'^\dagger)^m H_2 (V')^m H_2] = d \cos(2 m \theta)\; ,
\end{align}
we can zoom in on the previous estimation by applying   $V'$,  $m>1$ times,
and using the results of Sec.~\ref{blackbox}.
In this case,
$m$ coherent transports of the probe between Alice and Bob are required.

Notice that in this estimation procedure, the action
of relevant operations need not be controlled by the ancilla.
Finally, the other Euler angles can be estimated in a similar way if Alice
and Bob agree to apply other analogous operations.

%%%%%%%%%%%%%%%%%%%%%%%%%%%%%%%%%%%%%%%%%%
%%%%%%%%%%%%%%%%%%%%%%%%%%%%%%%%%%%%%%%%%%
\section{Parameter estimation vs. Grover's search algorithm in DQC1}
\label{grover}
So far we have shown that to reach certain precision in the
estimation of a parameter,  DQC1 requires less resources than other methods. 
One may wonder if
such a quantum speed-up can also be attained in problems such as
searching for a particular property in a given set (i.e., search
problem), which is the case for pure-state quantum
algorithms. However, in Ref.~\cite{KL98} the authors proved that
DQC1 is strictly less powerful than standard quantum computation in
the oracle setting. This implies that DQC1 does not provide a
quadratic quantum speed-up in the search problem as the one given by
Grover's algorithm~\cite{Gro96}. To show this, consider the situation
in which we are given a black box that implements either the unitary
$U_{B}=e^{i \theta \ket{S}\bra{S}}$, with $\theta \ne 0$, or
$U_{B}=\one$, over the state of the $n$ qubits in the probe. Here,
$\ket{S}$ encodes the solution to our search problem. We want to
determine the existence of a solution; that is, we want to specify if
the action of $U_{B}$ is trivial (i.e., $\one$) or not (i.e., $ e^{i
  \theta \ket{S}\bra{S}}$).  In fact, this is a phase
estimation problem in which we have to decide whether the phase is $0$
or $\theta$.

In general, the output of a DQC1 algorithm is given by
\begin{equation}
\langle \sigma_z^{\sf a} \rangle = \tr [ \rho_f \sigma_z^{\sf a}]\;,
\end{equation}
with $\rho_f$ the ancilla-probe state right before the measurement. With no loss of generality, 
\begin{equation}
\rho_f = W_Q U_{B}  \cdots W_1 U_{B} W_0 \rho_0 W^\dagger_0 U^\dagger_{B} W^\dagger_1 \cdots U^\dagger_{B} W^\dagger_Q\;,
\end{equation}
with $\rho_0 = (\ketbra{0}{\sf a} \otimes \one_n)/2^n$  being the initial state, and $Q$ the number of calls to $U_{B}$. Since $ \ketbra{0}{\sf a}=(\one_{\sf a}-\sigma_z^{\sf a})/2$, we have
\begin{equation}
\langle \sigma_z^{\sf a} \rangle = \tr[W_Q U_{B}  \cdots  U_{B} W_0 \sigma_z^{\sf a} W^\dagger_0 U^\dagger_{B} \cdots U^\dagger_{B} W^\dagger Q \sigma_z^{\sf a}]/2^{n+1}\;.
\end{equation}
Following the proof in Ref.~\cite{KL98}, we obtain
\begin{equation}
\left |  \langle \sigma_z^{\sf a} \rangle |_{U_B=\one} -  \langle \sigma_z^{\sf a} \rangle |_{U_B=e^{i \theta \ket{S}\bra{S}} } \right| \le 4Q/2^{n+1}\;.
\end{equation}
Since DQC1 returns $\langle \sigma_z^{\sf a}\rangle$ at  accuracy $\Delta$,  $Q$ must be exponentially large in $n$ or the algorithm needs to be executed exponentially many times to determine whether there is a solution or not. That is, if $J$ is the amount of times that the algorithm is executed, we would expect that the precision in the estimation  scales as $\Delta' = \cO(\Delta/\sqrt{J})$. To solve the problem, it is necessary (but not sufficient) to choose $J$ and $Q$ such that
\begin{equation}
4Q/2^{n+1} > \Delta' \;,
\end{equation}
requiring $N \equiv JQ$ uses of the black box. Thus,
\begin{equation}
\label{Nres}
N > \cO(2^n)\;.
\end{equation}
If $Q=\cO(\sqrt{2^n})$ as in Grover's algorithm, we need
$J=\cO(\sqrt{2^n})$ to satisfy Eq.~(\ref{Nres}), and no quantum
speed-up with respect to the classical counterpart is obtained in this
case.

The reason for the existence of a quantum-speed up in parameter
estimation is that  the unitary operators considered
act non-trivially in a large-dimensional subspace of the
corresponding $2^n$-dimensional Hilbert space. In this case, the
output signal obtained by executing the DQC1 circuits enables us to
distinguish between different unitaries and to estimate the parameter.
In the search problem, however, the operator $U_B$ is very close to
the identity operator $\one$ in that it only affects the state
$\ket{S}$.  Therefore, its action over highly-mixed states is almost
trivial. This is not the case in pure-state algorithms where one
usually works in a two-dimensional Hilbert space spanned by the states
$\ket{S}$ and $\ket{S^\perp}$, with
$\bra{S}S^\perp\rangle=0$~\cite{KOS07,Gro96}.

\vspace{1cm}

%%%%%%%%%%%%%%%%%%%%%%%%%%%%%%%%%%%%%
%%%%%%%%%%%%%%%%%%%%%%%%%%%%%%%%%%%%%
\section{Conclusions}
\label{concl}
Parameter estimation at the quantum metrology limit could have a wide
range of applications in metrology~\cite{GSM04, BIW96}. Further,
single-parameter estimation is related to phase estimation, a
cornerstone of quantum computation. Mixed-state quantum computation,
as formalized in the DQC1 model, is interesting both from a
theoretical and from a practical point
view~\cite{KL98,LKC02,VARIOUS2}. We have shown that, under fairly
general conditions, it is possible to perform parameter estimation at
the quantum metrology limit within the DQC1 model. These conditions
are presented using Lie algebraic methods. The algorithm proceeds
using adaptive Bayesian estimation. In each step we zoom in on the
previously estimated parameter, while ensuring that the increased
variance remains below certain bounds. A measurement reduces such a
variance to the previous value and this procedure is repeated until
the desired precision in the estimation is reached.  In sort, the
procedure ensures that the phase is kept in the same region, with
almost constant variance, but increasing winding number.

The adaptive estimation is clearer when the time of the
evolution under the unknown Hamiltonian (parameter) can be controlled at
will. When lacking this freedom, as in the case of
black box estimation, the algorithm for continuous time can be amended with some
straightforward modifications. Moreover, to perform multi-parameter estimation,
implementation of dynamical-decoupling techniques reduce the problem 
to several single-parameter estimation procedures. Yet,
these techniques are not necessary in some simple cases like frame
alignment between two parties, Alice and Bob. For spatial frame alignment,
the Euler angles can be estimated at the quantum metrology limit
with  Bob having access to the completely mixed, separable,  state of the probe only.
Surprisingly, although precise estimation is intimately
related to the quantum speed-up given by Grover's search algorithm, 
the later cannot be performed efficiently within the DQC1 model.

\acknowledgments We are thankful to E. Bagan, H. Barnum, C.M. Caves,
E. Knill, and L. Viola for interesting discussions. This work was
carried out under the auspices of the National Nuclear Security
Administration of the US Department of Energy at Los Alamos National
Laboratory under Contract No. DE-AC52-06NA25396 and partially
supported by ONR Grant No. N00014-07-1-0304,.

\appendix
\section{Lie algebra representations}
\label{ap:lie}
For $X(T)=W^\dagger(T) X W(T)$ we can also write $X(T) = X+iT [H,X]-T^2/2
[H,[H,X]]+ \cdots$, with $[Y,X]=YX-XY$.  That is, $X(T)$ is a linear
combination of observables belonging to the Lie algebra $\fh$
generated by $H$ and $X$~\cite{Cor97}. In general, $\fh \equiv
\{O_1,\ldots,O_M \}$ is an $M$-dimensional (real) semisimple Lie
algebra, with $O_j = O^\dagger_j$ and $\tr[O_iO_j]=d \delta_{ij}$
($d\in \mathbb{R}$). A faithful representation of $\fh$ is the mapping
$O_j \rightarrow \bar{O}_j$, with $\bar{O}_j= \bar{O}^\dagger_j$ being
$(m\times m)-$dimensional matrices that satisfy
\begin{eqnarray}
[O_i,O_j] = \sum_k f^k_{ij} O_k &\leftrightarrow& [\bar{O}_i,\bar{O}_j] = \sum_k f^k_{ij} \bar{O}_k \;, \\
\tr[\bar{O}_i \bar{O}_j] & =& \bar{d} \delta_{ij}\;,
\end{eqnarray}
for some $\bar{d} \in \mathbb{R}$. The coefficients $f^k_{ij}$ are the
so-called structure constants of $\fh$. Then, $X(T) = \sum_k c^k O_k$,
$c^k \in \mathbb{R}$, and 
\begin{equation}
\label{represent}
\bar{X}(T)=\bar{W}^\dagger(T) \bar{X} \bar{W}(T) = \sum_k c^k \bar{O}_k\;,
\end{equation}
with $\bar{W}(T) = e^{-i \bar{H}T}$. The $c^k$'s depend only on
the $f^k_{ij}$'s. Equation~\eqref{represent} implies that if two
different sets of (linearly-independent) matrices have the same
commutation relations, the coefficients $c^k$ can be determined by
working in either matrix representation:
\begin{equation}
c^k=\tr[X(T) O_k]/d=\tr[\bar{X}(T) \bar{O}_k]/\bar{d}\;.
\end{equation}

Because $\{H_0,H_1,H_2\}$ span a $\fsu(2)$ Lie algebra, they
satisfy
\begin{equation}\label{eq:2ie}
[H_j, H_k ] = 2 i \epsilon_{jkl} H_l\;,
\end{equation}
with $j,k,l \in \{0,1,2 \}$ and $\epsilon_{jkl}$ the totally
antisymmetric symbol. The $\fsu(2)$ Lie algebra can be built upon
$(2\times2)-$dimensional Hermitian, traceless, matrices (i.e., Pauli
spin-1/2 operators). This allow us to carry out the
calculation of Eq.~\eqref{represent} in a low-dimensional representation.

To give an example where the $L^2$ trace estimation can be reduced
[Eqs.~(\ref{coslc},\ref{sinlc})], consider again the situation where $H =
\theta \sum_{j=1}^n \sigma_z^j$ [Eq.~\eqref{mfield}]. Then, for
example,
\begin{equation}
\sigma_x^1(T) = e^{i H T} \sigma_x^1 e^{-i H T} = \cos(2 \theta T) \sigma_x^1 - \sin(2 \theta T) \sigma_y^1\;,
\end{equation}
as $[\sigma_x^1, \sigma_z^j ] = 0 \ \forall \ j>1$. That is, $\cos(2
\theta T)$ can be estimated by computing the renormalized trace of the unitary
$U=\sigma_x^1(T) \sigma_x^1 $ only. This situation can be generalized to the case when
$H_0=\sum_{\mu=1}^L e^{\mu,0} \sigma_{\mu,0}$, if there is an $H_1= \sigma_{\mathfrak 1}$, with $\sigma_{\mathfrak 1}$  a Pauli product such
that (for some $\mu$)
\begin{eqnarray}
\nonumber
  \left[ \sigma_{\mu,0}, \sigma_{\mu',0} \right] & = & 0 \quad \forall \ \mu,\mu'\;, \\
  \nonumber
  \left\{ \sigma_{\mu,0} , \sigma_{\mathfrak 1} \right\} & = & 0 \;, \\
  \left[ \sigma_{\mu',0}, \sigma_{\mathfrak 1}  \right] &=& 0  \quad \forall \ \mu' \ne \mu\;.
\end{eqnarray}
It follows that
\begin{eqnarray}
\label{comm5}
\nonumber
  [\sigma_{\mu,0},\sigma_{\mathfrak 1}]&=&2 i \sigma_{\mathfrak 2}\;, \\
  \label{comm6}
  \left[ \sigma_{\mu',0}, \sigma_{\mathfrak 2}  \right] &=& 0  \quad \forall \ \mu' \ne \mu\;,
\end{eqnarray}
with $\sigma_{\mathfrak 2}=-i(\sigma_{\mu,0}\sigma_{\mathfrak 1})$.
Equation~(\ref{comm6}) results from the Jacobi identity and the
definition of $\sigma_2$. Thus, $\{\sigma_{\mu,0}, \sigma_{\mathfrak
  1}, \sigma_{\mathfrak 2} \}$ satisfies the $\fsu(2)$-commutation
relations, yielding
\begin{eqnarray}
\label{traceest3}
\cos(2 \theta e^{\mu,0} T) &=& \tr[W^\dagger(T) \sigma_{\mathfrak 1} W(T) \sigma_{\mathfrak 1} ]/2^n\;, \\
\label{traceest4}
\sin(2 \theta e^{\mu,0} T) &=& \tr[W^\dagger(T) \sigma_{\mathfrak 1} W(T) \sigma_{\mathfrak 2} ]/2^n\;.
\end{eqnarray}
We estimate Eqs.~(\ref{traceest3}) and~(\ref{traceest4}) 
by executing the circuit of Fig.~\ref{dqc1fig} with
$U=W^\dagger(T)\sigma_{\mathfrak 1} W(T) \sigma_{\mathfrak 1}$ and $U=W^\dagger(T)\sigma_{\mathfrak 1} W(T) \sigma_{\mathfrak 2}$, respectively.

\section{The adaptive Bayesian estimation procedure}\label{ap:bayes}

To reach the QML in our estimation procedure, we first assume an
initial estimation of  $\theta$ given by
a prior distribution $\cN (\hat \theta_0, \Delta ')$.  For
simplicity, in the following we focus on the case where $\theta$ is
determined from the estimation of $\cos(2 \theta T)$ for different
values of $T$, denoted as $T_l$.  This is done using the
DQC1 algorithm of Fig.~\ref{dqc1fig3}, where the corresponding Pauli
products $\sigma_{\mu,j}$ are determined by Eq.~(\ref{coslc}).  Thus,
$x_l$ denotes an estimate of $\cos(2 \theta T_l)$ and we consider $
\pi>\hat \theta_0 > 0$.  Otherwise, the estimation of $\sin(2 \theta
T_l)$ is also required to determine, for example, the
corresponding quadrant of $\theta$.

To obtain $x_1$, we choose $T_1$ such that $2 \hat \theta_0 T_1 =
\pi/2$~\cite{comment1}.  This can be done with an upper bounded
initial use of resources when $\hat \theta_0 \in [\chi,\pi)$, with
 $\chi>0$.  That is, $ 1/4 < T_1 \le \pi/(4\chi)$.
Nevertheless, if $\hat \theta_0 \ll 1$ (i.e., $T_1 \gg 1$), a similar
analysis as the one carried out below can be performed by measuring
$\sin(2 \theta T'_1)$ instead of $\cos(2 \theta T_1)$, with $T'_1
=\cO( 1)$~\cite{comment2}.  Therefore, we take $\cN(\pi/2,(c\, \Delta)^2)$
as the prior distribution of $\alpha_1= 2 \theta T_1$.  We define here
$\Delta$ to be the output precision of DQC1 when measuring $\cos(2
\theta T)$, which may actually involve many ($L^2>1$) different runs
of the circuit of Fig.~\ref{dqc1fig3}.  
We assume $1 \gg c\Delta \gg \Delta$.  The
measurement outcome $x_1$ of the first measurement has then a sampling
distribution given by $\cN (\cos(\alpha_1),
\Delta ^2)$.

The joint distribution $f(x_1, \alpha_1) = f(x_1|\alpha_1)
f(\alpha_1)$ is
\begin{align}
\label{dens1}
f( x_1, \alpha_1) &=  \frac 1 { \sqrt{2 \pi} \Delta } e^{-\frac {(x_1 - \cos(\alpha_1))^2}
{2 \Delta^2}} \frac 1 { \sqrt{2 \pi} c \Delta } e^{-\frac {(\alpha_1 - \pi/2)^2}{2 c^2 \Delta^2}} 
 \nonumber \\
& \approx \frac 1 { \sqrt{2 \pi} \Delta } e^{-\frac {(x_1 + \alpha'_1
)^2}{2 \Delta^2}} \frac 1 { \sqrt{2 \pi} c \Delta
 } e^{-\frac {( \alpha'_1)^2}{2 c^2 \Delta^2}}\; .
\end{align}
Here, $\alpha'_1 = \alpha_1 - \pi/2$ and, for simplicity,  we approximated
$\cos(\alpha_1)$ at first order by $-\alpha'_1$ so that the joint distribution
is normal. The error in this
approximation can be bounded above, with high confidence, as
\begin{equation}
\label{coserror}
 |\cos(\alpha_1) - (-\alpha'_1)| \le \delta = (c'\Delta)^3 /6 \ll 1 \;,
\end{equation}
for some $c' \ge c$. For example, if choosing $c'=1.96 c$,  Eq.~\eqref{coserror} 
determines a $95\%$ credible interval for $\cos(\alpha_1)$. Such a confidence
can be made exponentially close to 1 as $c'$ increases. 
Of course this error can be avoided if no approximation is made and other
analytical or numerical methods are used. 
Nevertheless, a linear approximation to
the cosine is enough for our purposes, as it will yield the proper
results~\cite{comment2b}. Moreover, the  error of the above approximation will be
further corrected by subsequent measurements. This is a consequence
of the adaptive method.

The following step is to update the information about $\alpha_1$ (or
$\theta$) based on the measurement outcome $x_1$. The posterior
distribution is $f(\alpha_1 | x_1) = f(x_1,\alpha_1)/f(x_1)$.
Using Eq.~(\ref{dens1}) this distribution can be shown to be
\begin{align}
\label{dens2} 
f(\alpha_1|x_1)\approx \cN\left(\hat{\alpha}_1 , \Delta_1^2\right)\; .
\end{align}
$\hat{\alpha}_1$ and $\Delta_1$ are determined from the exponent
$E=-(x_1 + \alpha'_1)^2 / (2\Delta^2) - \alpha'^2_1/(2c^2 \Delta^2)$
in Eq.~(\ref{dens1}).  We write
\begin{equation}
\label{dens3}
E= - \frac{\left( \alpha'_1 + \frac{c^2}{1+c^2} x_1  \right) ^2}{ 2 \Delta_1^2} + g(x_1)\;,
\end{equation}
implying
\begin{eqnarray}
\label{dens4a}
\Delta_1 &=& \frac{c}{\sqrt{1+c^2}} \Delta < \Delta\;, \\
\label{dens4}
\hat{\alpha}_1 &=& \pi/2 - \frac{c^2}{1+c^2} x_1 \;.
\end{eqnarray}
Summarizing, if $\hat \theta_1$ is our estimator of $\theta$ after the
first measurement, we have
\begin{equation}
\label{estim1}
\hat{\theta}_1 = \frac {\hat \alpha_1}{2 T_1} = \frac{1}{2T_1} \left( \pi/2 - \frac{c^2}{1+c^2}x_1 \right)\;, 
\end{equation}
which is only a linear correction in $x_1$.
Moreover, our knowledge about $\theta$ has increased such that
\begin{equation}
\label{error1}
\hat\theta_1 - 1.96 \Delta_1/ (2 T_1) \le \theta \le \hat \theta_1 +1.96 \Delta_1/ (2 T_1)
\end{equation}
is a $95\%$ credible interval. If the desired output precision
$\Delta_\theta$ in the single-parameter estimation satisfies $
\Delta_1/(2T_1) \le \Delta_\theta$, the estimation procedure stops
here.  Otherwise, further measurements are required.

In the previous analysis we have neglected other values of $\alpha_1
\mod 2\pi$ that would yield the same measurement outcome. This
assumption introduces an extremely small error bounded above by
\begin{equation}
\text{erfc}\left( \frac{2 \pi}{c \Delta \sqrt{2}} \right)\;,
\end{equation}
where erfc denotes the {\em complementary error function} of the
normal distribution.  Since $c\Delta \ll 1$, we have $\text{erfc}( 2
\pi/ (c \Delta \sqrt{2})) \approx 0$.  For example, for the
unrealistic case of $c\Delta = 1$, we have $\text{erfc}( \sqrt{2} \pi)
\approx 3.4 \ 10^{-10}$. Thus, our assumption does not affect the
final results of the estimation procedure.

To increase the precision we zoom in on the previously estimated $
\theta$, so a better estimate is attained. Since $x_1 = \cO(c \Delta)
\ll 1$~\cite{comment3}, we have $|2 \hat \theta_1 T_1 - \pi/2| \ll 1$.
Accordingly, there exists a time period $T_2 > T_1$ such that
\begin{equation}
\label{loops}
2 \hat \theta_1 T_2 = \pi/2 +2 p_2 \pi\;,
\end{equation}
with $p_2 \in \mathbb{N}^*$. $T_2$ denotes the evolution time in the
second measurement (estimation).  The main reason behind
Eq.~(\ref{loops}) is that here, as in the first measurement, we seek to
make a phase estimation around $\pi/2 \mod 2\pi$. In this region, the
cosine function is more sensitive to variations in the phase.  The
second measurement returns $x_2$, an estimate of $\cos(\alpha_2)$,
with $\alpha_2 = 2 \theta T_2$.

Since $T_2 = a_2 T_1$, with $a_2 >1$, we obtain $a_2 \approx (1+4p_2)$
[see Eqs.~\eqref{estim1} and \eqref{loops}].  The previously estimated
$\theta$ has a variance of order $\Delta/(2 T_1)$ [Eqs.~\eqref{dens4a}
and~\eqref{error1}], so the variance of $\alpha_2$ is of order $a_2
\Delta$.  To guarantee that this variance is similar to that of
$\alpha_1$ (first measurement), we choose $a_2$ as large as possible so
that~\cite{comment4}
\begin{align}
  \label{eq:offset}
  0 < c'- 4< a_2 \le c'\;.
\end{align}
This implies that Eq.~\eqref{coserror} is still satisfied when
replacing $\alpha_1$ by $\alpha_2$, and $\alpha'_1$ by $\alpha'_2 =
\alpha_2 - (\pi/2 + 2 p_2 \pi) $.  As an example, consider the case
when $c' \approx 10$.  Therefore, we choose $a_2 \approx 9 \le c'$,
corresponding to $p_2=2$ in Eq.~(\ref{loops}).

After the measurement, the outcome $x_2$ is used to update our
information about $\theta$. Since $c' \Delta \ll 1$, $\alpha_2 \in
[2p_2 \pi, (2p_2+1) \pi]$ with large confidence. That is, we do not
consider other values of $\alpha_2 \mod 2\pi$ and we make the
estimation in this region only.  The joint distribution is now
\begin{align}
\label{dens5}
f( x_2, &\alpha_2| x_1) = f(x_2|x_1, \alpha_2) f(\alpha_2| x_1) \nonumber \\
& \approx \frac 1 { \sqrt{2 \pi} \Delta } e^{-\frac {(x_2 + \alpha'_2
    )^2}{2 \Delta^2}} \frac 1 { \sqrt{2 \pi} a_2 \Delta_1 } e^{-\frac
  {(\alpha'_2 )^2}{2 (a_2)^2 \Delta_1^2}}\;.
\end{align}
$f(\alpha_2|x_1)$ has been determined using
Eq.~(\ref{loops}). Thus, the posterior density distribution
$f(\alpha_2|x_2,x_1) = f( x_2, \alpha_2| x_1) / f(x_2 | x_1)$ is a
normal $\cN(\hat{\alpha}_2, (\Delta_2)^{2})$, where $\hat{\alpha}_2$
and $\Delta_2$ are determined by the exponent appearing in
Eq.~(\ref{dens5}). These are
\begin{eqnarray}
\hat{\alpha}_2 &=& (\pi/2 + 2p_2\pi) -  \frac{(a'_2)^2}{1+(a'_2)^2} x_2, \\
\label{dens6}
\Delta_2 &=& \frac{a'_2}{\sqrt{1+(a'_2)^2}} \Delta < \Delta\;,
\end{eqnarray}
with
\begin{align}
  a'_2 = a_2 \Delta_1/\Delta < a_2 \;.
\end{align}
Summarizing, the estimator after the second measurement is
\begin{equation}
\hat{\theta}_2 = \frac{1}{2T_2} \left( (\pi/2+2p_2\pi) -  \frac{(a'_2)^2}{1+(a'_2)^2} x_2 \right)\;,
\end{equation}
with a $95\%$ credible interval
\begin{equation}
\label{error2}
\hat\theta_2 - 1.96 \Delta_2/(2T_2) \le \theta \le \hat \theta_2 + 1.96 \Delta_2/(2T_2) \;.
\end{equation}
The standard deviation in the estimation of $\theta$ has
been reduced by a factor of order $a_2$ with respect to the one
returned in the first measurement [Eq.~(\ref{error1})].

If the desired output precision satisfies $ \Delta_2/(2T_2) \le
\Delta_\theta $, the estimation stops here. Otherwise, we continue
with the adaptive procedure. At each step $l$ we find $T_l$ such that
\begin{equation}
2 \hat \theta_{l-1} T_l =  \pi/2 + 2p_l \pi\;,
\end{equation}
where $\hat \theta_{l-1}$ is the Bayes' estimator determined by the
previous measurement outcomes. Since $p_l > p_{l-1}$ are positive
integers, we write $T_l = a_l T_{l-1}$, with $a_l \approx
(1+4p_l)/(1+4 p_{l-1})$.  The $l$th measurement returns $x_l$, an
estimate of $\cos(\alpha_l)$, with $\alpha_l = 2 \theta T_l$. This is
done by running the algorithm of Fig.~\ref{dqc1fig3} with $T=T_l$. To
keep the variance of $\alpha_l$ at order $c' \Delta$ we choose $a_l$
such that $c' - 4 \le a_l \le c'$~\cite{comment4}. With this choice,
Eq.~\eqref{coserror} is still satisfied when replacing $\alpha_1$ by
$\alpha_l$, and $\alpha'_1$ by $\alpha'_l = \alpha_l - (\pi/2 + 2 p_l
\pi)$.  
Using Bayes'
rule, and considering
\begin{align}
\label{dens8ap}
f( x_l, &\alpha_l| x_{l-1},\ldots, x_1) \approx \nonumber \\ &\approx
\frac 1 { \sqrt{2 \pi} \Delta } e^{-\frac {(x_l + \alpha'_l )^2}{2
    \Delta^2}} \frac 1 { \sqrt{2 \pi} a_l \Delta_{l-1} } e^{-\frac
  {(\alpha'_l )^2}{2 (a_l)^2 \Delta_{l-1}^2}}\;,
\end{align}
we obtain for the $l$th estimation
\begin{eqnarray}
\label{dens7a}
2 \hat \theta_{l} T_{l} &=& \pi/2 + 2p_{l}\pi -  \frac{(a'_{l})^2}{1+(a'_{l})^2} x_{l} \;, \\
\label{dens7}
\frac {\Delta_{l}} {2T_{l}} &=& \frac{a'_{l}}{\sqrt{1+(a'_{l})^2}} \frac{\Delta}{2T_{l}} < \frac {\Delta}{2T_{l}}\;,
\end{eqnarray}
with
\begin{align}
  a'_{l} = a_{l} \Delta_{l} /\Delta < a_{l}\;.
\end{align}
That is, the variance of the $l$th estimator has been reduced by
a factor $1/T_l$.

We now show that the QML has been achieved. Consider the total
evolution time $T_t=T_1 + \cdots + T_K $, for $K$ estimations, with $T_l = (\prod_{l'=2}^l
a_{l'} )T_1$.  Then,
\begin{equation}
T_t =\left[ \frac{1}{a_2 \ldots a_K} + \ldots + 1 \right] T_K\;.
\end{equation}
Moreover, since $(1/a_l) \le 1/(c'-4) <1$, we have
\begin{equation}
\label{ttime}
T_t \le [(c'-4)^{-(K-1)} + \cdots +1] T_K < \frac{c'-4}{c'-5} T_K\;.
\end{equation}
Equation~\eqref{ttime} gives the total resource scaling $T_t =
\cO[\Delta/(2 \Delta_\theta)]$ (i.e., the total evolution time under
the action of $H$), implying that the QML is attained.

\section{First step in black-box estimation}\label{ap:discrete}
For simplicity, assume $\pi/4 > \hat \theta_0 > 0$, with $\hat \theta_0$ the mean of
the prior (normal) distribution of $\theta$. Then, there exists a $\phi_1$ such that
\begin{equation}
2 \hat \theta_0 + 2 \phi_1 = \pi/2\;.
\end{equation}
The first estimation of $\theta$ can be done by running the algorithm
of Fig.~\eqref{dqc1fig3}  replacing $W(T) \rightarrow
W_a(1,\phi_1) = W_B e^{-i \phi_1 H_0}$. Since $H_0$ is known,
$W_a(1,\phi_1)$ can be implemented with available gates.  The output
of the first measurement, denoted by $y_1$, is a measurement of
$\cos(\beta_1)$, with $\beta_1 = 2 (\theta +\phi_1)$. The a priori
distribution of $\beta_1$ is then given by $f(\beta_1) =
\cN(\pi/2,(c\Delta)^2)$~\cite{comment6}. Similar to the continuous time evolution
case, this distribution yields the joint distribution
\begin{align}
\label{bbdens1}
f( y_1, \beta_1) &=  \frac 1 { \sqrt{2 \pi} \Delta } e^{-\frac {(y_1 - \cos(\beta_1))^2}
{2 \Delta^2}} \frac 1 { \sqrt{2 \pi} c \Delta } e^{-\frac {(\beta_1 - \pi/2)^2}{2 c^2 \Delta^2}} 
 \nonumber \\
& \approx \frac 1 { \sqrt{2 \pi} \Delta } e^{-\frac {(y_1 + \beta'_1
)^2}{2 \Delta^2}} \frac 1 { \sqrt{2 \pi} c \Delta
 } e^{-\frac {( \beta'_1)^2}{2 c^2 \Delta^2}}\; ,
\end{align}
with $\beta'_1= \beta_1 - \pi/2$. The first measurement returns an estimator of $\beta_1$, with mean $\hat{\beta}_1= \pi/2 - [c^2/(1+c^2)]y_1$, and  standard deviation
$\Sigma_1=(c/\sqrt{1+c^2}) \Delta < \Delta$. Therefore, the first estimation of $\theta$ has mean and variance determined by
\begin{align}
\label{bbmean1}
\hat \theta_1 = \frac{1}{2} \left( \pi/2 -\frac{c^2}{1+c^2}y_1 \right) - \phi_1 \; , \\
\label{bbvar1}
\Sigma_1'=\Sigma_1/2 = \frac{c}{\sqrt{1+c^2}} \frac{\Delta}{2} < \Delta/2\;.
\end{align}
Clearly, the accuracy of the estimation has increased after the first measurement. Thus,  we continue with the adaptive Bayesian method by zooming in on the previously estimated parameters.


\begin{thebibliography}{12}
\bibitem{comment-0} If, on the contrary, the interaction determines an
  energy range that scales like $n^k$, the QML precision achieved by
  these protocols is $\cO(1/n^kT)$~\cite{BFCG09}.
\bibitem{comment-1b}
On the contrary, for fixed $T=1$, $\Delta_\theta = \cO(1/\sqrt{n})$ so 
nothing is gained with respect to the SQL in this case.
\bibitem{comment0} Such a decomposition is always possible because
  Pauli products constitute a basis of $\fsu(2^n)$.
\bibitem{comment8} All of the above assumptions
are satisfied in the first measurement, as seen in Appendix~\ref{ap:discrete}.
\bibitem{comment7} To obtain $\Delta_\gamma$ we have assumed first
  that $c \Delta$ is the standard deviation associated with the prior
  distribution of every $\theta^\nu$. Since
  $|\gamma(\epsilon,p,T_l,\overrightarrow \theta)| \le \varepsilon$,
  it is expected that $\Delta_\gamma < c\Delta \epsilon$.  Moreover,
  since $\gamma(\epsilon,p,T_l,\overrightarrow \theta)$ depends on
  many stochastic variables, we also assumed that $f(\gamma)$ is
  unbiased.  This assumption may certainly not be right in a general
  scenario and we would expect that, for some simple cases, such a
  bias could be calculated and corrected. However, the error
  introduced by approximating the mean of
  $\gamma(\epsilon,p,T_l,\overrightarrow \theta)$ by $0$ is also of
  order $\varepsilon$, with negligible effects in our calculation.
\bibitem{comment1} 
We have chosen $\pi/2$ because the cosine function in this region
is very sensitive to variations in the phase. Nevertheless, 
 our analysis can still be performed choosing another $T_1$, as
 long as $\sin(2 \hat \theta_0 T_1)$ is not very close to 0.
\bibitem{comment2} 
This can be done by choosing the Pauli products of
  Eq.~\eqref{sinlc} in the algorithm of Fig.~\ref{dqc1fig3}, or by
  compensating the phase by $(\pi/2)-\hat\theta_0$ in the first
  measurement.  Such a phase compensation will be modified as we learn
  about $\theta$ (see Sec.~\ref{blackbox}).
\bibitem{comment2b} 
Numerical and asymptotic analyses show that the linear approximation results in an error $\cO[(c \Delta/T_1)^2] \ll \Delta/T_1$ in the estimator of Eq.~\eqref{estim1} and an error of $\cO[(c \Delta)^4]$ in the computed standard deviation.
\bibitem{comment3}
It is expected that the first measurement outcome lies, in the worst case scenario, within 
a few standard deviations $c \Delta$ from $0$.
\bibitem{comment4} The main necessary condition for choosing $a_2$ is
  that the different phases $\mod 2 \pi$ do not get aliased when
  measuring $\cos(2 \theta T_2)$. Nevertheless, we require $a_2 \Delta
  \ll 1$, so a linear approximation to the cosine function is still
  possible. We have chosen, for simplicity, $a_2 = \cO(c')$. The
  offset of $4$ in Eq.~\eqref{eq:offset} is due to $a_2 \approx
  (1+4p_2)$ and $p_2 \in \mathbb{N}^*$.
\bibitem{comment6} The constant $c$ in this case may be even smaller
  than the one introduced in Sec.~\ref{ap:bayes}.  The normal
  distribution is due to $\beta_1$ being a linear transformation of
  $\theta$.
\end{thebibliography}
\end{document}